\documentclass{aa}
\usepackage{graphicx}
\usepackage{txfonts}
\renewcommand{\ion}[2]{#1\,{\sc #2}}
\newcommand{\lya}{Ly$\alpha$}

\begin{document}
   \title{New observations of the extended hydrogen exosphere\\of the extrasolar planet HD\,209458b
            \thanks{Based on observations made with the Advanced Camera for Surveys on board the \emph{Hubble Space
            Telescope}.}
   }

   \author{
   D.~Ehrenreich\inst{1} \and A.~Lecavelier des Etangs\inst{2} \and G.~H\'ebrard\inst{2} \and J.-M.~D\'esert\inst{2} \and A.~Vidal-Madjar\inst{2}\and \\
   J.~C.~McConnell\inst{3} \and C.~D.~Parkinson\inst{4} \and G.~E.~Ballester\inst{5}
   }

   \offprints{D.\ Ehrenreich}

   \institute{
   Laboratoire d'astrophysique de l'observatoire de Grenoble, Universit\'e Joseph Fourier, CNRS (UMR 5571)\\
   BP 53, 38041 Grenoble cedex 9, France\\
   \email{david.ehrenreich@obs.ujf-grenoble.fr}
    \and
   Institut d'astrophysique de Paris, Universit\'e Pierre \& Marie Curie, CNRS (UMR 7095)\\
   98 bis, boulevard Arago 75014 Paris, France
    \and
   Department of Earth and Space Science and Engineering, York University\\
   4700 Keele street, Toronto, ON M3J1P3, Canada
    \and
   Department of Atmospheric, Oceanic, and Space Sciences, University of Michigan\\
   2455 Hayward street, Ann Arbor, MI 48109, USA
    \and
   Lunar and Planetary Laboratory, University of Arizona\\
   1040 East 4th street, Tucson, AZ 85721-0077, USA
   }

   \date{Accepted in \emph{Astronomy \& Astrophysics}.}

  \abstract
    {Atomic hydrogen escaping from the planet \object{HD\,209458b} provides the largest
    observational signature ever detected for an extrasolar planet atmosphere.
    However, the Space Telescope Imaging Spectrograph (STIS) used in previous
    observational studies is no longer available, whereas additional observations
    are still needed to better constrain the mechanisms subtending the evaporation process,
    and determine the evaporation state of other `hot Jupiters'.}
    {Here, we aim to detect the extended hydrogen exosphere of HD\,209458b with
    the Advanced Camera for Surveys (ACS) on board the \emph{Hubble Space Telescope}
    (\emph{HST}) and to find evidence for a hydrogen comet-like tail trailing the
    planet, which size would depend on the escape rate and the amount of ionizing radiation emitted by the star.
    These observations also provide a benchmark for other transiting planets,
    in the frame of a comparative study of the evaporation state of close-in giant planets.}
    {Eight \emph{HST} orbits are used to observe two transits of HD\,209458b. Transit light curves are obtained
    by performing photometry of the unresolved stellar Lyman-$\alpha$ (\lya) emission line during
    both transits. Absorption signatures of exospheric hydrogen during the
    transit are compared to light curve models predicting a hydrogen tail.}
    {Transit depths of $(9.6\pm7.0)\%$ and $(5.3\pm10.0)\%$ are measured on the whole \lya\ line in visits~1 and~2, respectively.
    Averaging data from both visits, we find an absorption depth of $(8.0\pm5.7)\%$, in good agreement with previous studies.}
    {The extended size of the exosphere confirms that the planet is likely loosing hydrogen to space.
    Yet, the photometric precision achieved does not allow us to better constrain the hydrogen mass loss rate.}


    \authorrunning{D.\ Ehrenreich et al.}
    \titlerunning{Observations of HD\,209458b hydrogen exosphere}

   \maketitle

\section{Introduction}

Extrasolar planets transiting their parent stars represent $\sim10\%$ of the
total number of planets detected so far in the solar neighborhood (Schneider
2008). Planetary transits are rare and precious indeed, since they are powerful
tools to extract key physical properties of planets, like their mean densities,
chemical compositions, or atmospheric structures. In fact, during a transit,
the stellar light is partially filtered by the planetary atmosphere before
reaching the observer, who can thus probe the atmospheric limb structure and
composition with a differential spectroscopic analysis.

While detailed models of extrasolar planet transmission spectroscopy have been
flourishing (see, e.g., Seager \& Sasselov 2000; Brown 2001; Hubbard et al.\
2001; Ehrenreich et al.\ 2006; Tinetti et al.\ 2007), few detections of
atmospheric signatures have been reported. Most studies have been focused on
the giant `hot-Jupiter' HD\,209458b, the first extrasolar planet to be observed
in transit (Charbonneau et al.\ 2000; Henry et al.\ 2000). Atomic species have
been evidenced through the supplementary absorption they are triggering at key
wavelengths. Sodium (\ion{Na}{i}~D $\lambda$539~nm) in the lower atmosphere of
the planet gives rise to a $\sim10^{-4}$ extra-absorption (Charbonneau et al.\
2002). Hot hydrogen (\ion{H}{i}$^{*}$) in the middle atmosphere provides a
$\sim10^{-4}$ absorption around the Balmer jump (Ballester et al.\ 2007).
Compared to these rather tenuous signals, the $\sim10\%$ absorption signatures
of hydrogen (\ion{H}{i} $\lambda$121.6~nm), carbon (\ion{C}{ii}
$\lambda$133.5~nm), and oxygen (\ion{O}{i} $\lambda$130.2~nm) detected by
Vidal-Madjar et al.\ (2003, 2004) seem huge. In fact, those elements are seen
in the extended upper atmosphere of the planet, the exosphere.

The proximity of HD\,209458b to its star ($0.045$ astronomical units) makes the
gaseous planet receiving a colossal amount of extreme ultraviolet (EUV)
irradiation. Such an energetic input heats the hydrogen upper atmosphere to
$\sim 10^4$~K and inflates it to the Roche limit: the atmosphere is evaporating
and its spatial extent produces the large absorption observed in the
Lyman-$\alpha$ (\lya) stellar emission line (Vidal-Madjar et al.\ 2003).
Elements heavier than hydrogen (\ion{C}{ii} and \ion{O}{i}) are transported
upward by the hydrodynamic flow of escaping H atoms and consequently also give
rise to significantly large absorptions (Vidal-Madjar et al.\ 2004). The
evaporation process of the upper atmosphere has a solid theoretical ground on
which models (Lecavelier des Etangs et al.\ 2004; Yelle 2004; Tian et al.\
2005; Garc\'\i a-Mu\~noz 2007; Lecavelier des Etangs 2007; Penz et al.\ 2007)
are estimating the escape rate $\dot{M}$ of hydrogen and comparing it to
observational constraints. So far, theoretical and observational estimations of
$\dot{M}$ for HD\,209458b are converging (Yelle 2006).

Yet, important questions remain. Is there a hydrogen comet-like tail trailing
the planet? Is the size of this cloud fluctuating depending on the stellar
activity and variations in the flux of ionizing EUV radiation? What is the
evaporation state of other close-in planets? In fact, observational insights
from other systems would significantly constrain and refine the modeling of the
evaporation process in a more general frame. However, previous observations
were accomplished with the Space Telescope Imaging Spectrograph (STIS) on board
the \emph{Hubble Space Telescope} (\emph{HST}). The failure of STIS in August
2004 prevented the achievement of additional observations of exoplanets
transiting bright stars like \object{HD\,209458}, \object{HD\,189733}, or
\object{HD\,149026}.

In addition, Ben-Jaffel (2007) recently challenged the interpretation of
Vidal-Madjar et al.\ (2003) concerning the evaporation of HD\,209458b, on the
basis of a new analysis of the archival STIS data set. Although the apparent
disagreement between those authors has been clarified (Vidal-Madjar et al.\
2008), new observations of HD\,209458b are required to provide (i) additional
observational ground to the evaporation process and (ii) a benchmark
observation in the frame of a comparative study of the evaporation state of
hot-Jupiters.

These observations were performed with the Advanced Camera for Surveys (ACS;
Ford et al.\ 2003) on board \emph{HST}; this article describes their analysis.

\section{Observations of HD\,209458}

The observing program was originally designed to observe HD\,209458b with
\emph{HST}/STIS. After STIS failure, it has been possible to execute the
program with \emph{HST}/ACS during Cycle~13. The program (GO\#10145) consists
in $3+2$ visits, performed with the ACS/High Resolution Camera (HRC) and the
ACS/Solar Blind Camera (SBC), respectively. The first 3 visits, aimed at
characterizing the near-ultraviolet transmission spectrum of the planet, are
under on going analysis (D\'esert et al., in preparation). We focus here on the
last 2 visits, designed for a new observation of the evaporating hydrogen
envelope at \lya.

To this purpose, HD\,209458b was observed with the \emph{HST}/ACS/SBC on 2006
May 14 (visit~1) and 2006 May 31 (visit~2). Eight \emph{HST} orbits in total
were used for the following phase coverage of the transit light curve: 3
\emph{HST} orbits were obtained before 1st contact, 4 orbits between 1st and
4th contacts, and 1 orbit after 4th contact. Each \emph{HST} orbit consists in
8 exposures: a direct image of the star is first made with the F115LP filter; 6
exposures are then acquired in slitless spectroscopy mode with the PR110L
prism; another direct image is finally taken with the F115LP filter.

The data reduction described below makes use of the standard calibration
(flat-fielded and dark-subtracted) products of the ACS pipeline.

\section{Methods}

\begin{figure}
\resizebox{\hsize}{!}{\includegraphics{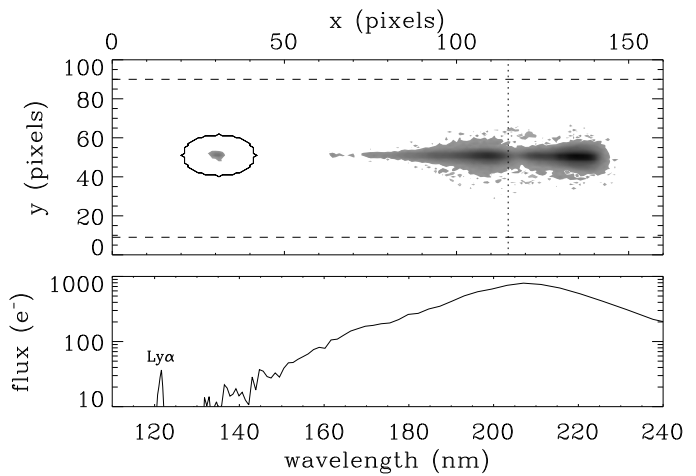}} \caption{Median spectrum of
HD\,209458b. The image in the upper panel is the median of all co-aligned prism
exposures. The dispersion direction is close to the detector $x$-axis. The
spectrum is unresolved along the $y$-axis. The background is estimated in upper
and lower part of the image (above and below the dashed lines). A circular
aperture of 10 pixels in radius in represented around the \lya\ emission line.
The image is stacked along the $y$-axis to produce the spectrum plotted in the
lower panel. The spectrum spans roughly from \lya\ to the dotted line in the
image at 240~nm, just after the continuum peak. The bright structure redward to
the continuum is an artifact due to the decrease of the resolution, the `red
pile-up'} \label{fig:spectrum_2d}
\end{figure}

Slitless spectroscopy yields spectral images similar to that represented in
Fig.~\ref{fig:spectrum_2d}. The direction of spectral dispersion is given by
the spectral trace, which can be linearly approximated for SBC data and make
with the detector $x$-axis an angle of $\sim 0.5\degr$. The spectral image has
three remarkable structures (from short to long wavelengths): the \lya\ stellar
emission line, the stellar continuum between 160 and 200~nm, and the `red
pile-up'. This last feature is an artifact resulting from the built up of
photons on a few detector pixels, as the spectral resolution significantly
diminishes toward the red. In fact, the spectral resolution of the SBC/PR110L
prism is $\lambda / \Delta\lambda \sim 250$ at \lya\ (2 pixels at
0.25~nm\,pix$^{-1}$), comparable to the resolution of Vidal-Madjar et al.\ 's
(2004) observations; it is $\sim40$ (2 pixels at 2.5~nm\,pix$^{-1}$) at 180~nm,
and becomes $\sim 10$ at 330~nm (2 pixels at 15~nm\,pix$^{-1}$).

Since the stellar continuum is very low in the ultraviolet -- typically lower
than the background level --, it is convenient to perform aperture photometry
on the unresolved \lya\ emission line. In fact, this allows us to make a more
direct use of the spectral images provided by the prism, aperture photometry
providing the total flux in the \lya\ line at a given time. Since the \lya\
line is unresolved, no spectral information can be extracted from these data.

The SBC detector is a multi-anode microchannel array (MAMA), i.e., a
photon-counting device. Hence, the flux at \lya\ is calculated for each image
by summing the pixels within a circular aperture of 10 pixels in radius
(represented in Fig.~\ref{fig:spectrum_2d}). We checked that the obtained
results do not significantly depend on the size of the photometric aperture for
aperture radii chosen between 5 and 20 pixels. Errors on the measured fluxes
are calculated from the tabulated error on each pixel in the aperture. Images
are previously co-aligned with a reference exposure; the offsets between the
spectra in two different images are determined by performing a
cross-correlation in Fourier space.

To estimate the background brightness, we used the 2D images (see
Fig.~\ref{fig:spectrum_2d}). On these images, we defined two bands above and
below the spectrum along the detector $x$-axis. These bands are 10 pixel-wide
in the $y$-direction. For each image at a given position $x$, the background is
calculated as the average of the measurements in the bands at the same $x$
position. Since the $x$-axis is close to the dispersion direction, these
estimations should therefore contain any wavelength-dependent structure of the
background. The background is fitted along the detector $x$-axis by a
4th-degree polynomial, which is then subtracted to each pixel row of the
detector image.

\begin{figure}
\resizebox{\hsize}{!}{\includegraphics{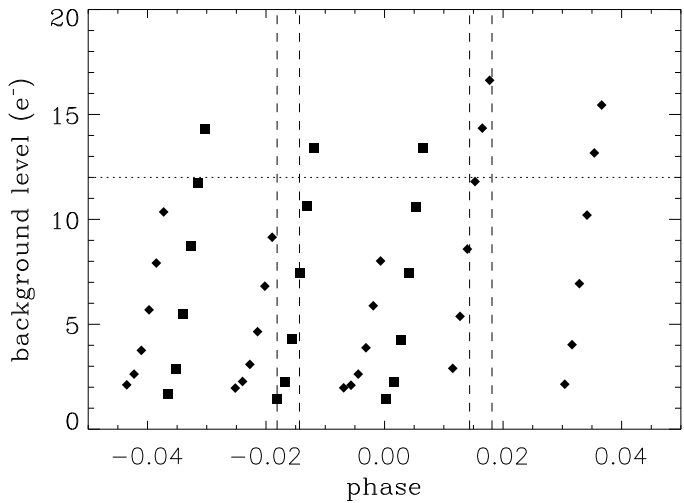}} \caption{Background level in
individual exposures of visit~1 (filled diamonds) and~2 (filled squares). The
median level of the \lya\ emission peak is indicated by the dotted line. Dashed
lines represent the four contacts of the transit.} \label{fig:background}
\end{figure}

The mean values of the background fits (averaged on the detector $x$-axis) are
represented as a function of the orbital phase in Fig.~\ref{fig:background}. A
clear rise of the background is seen along each orbit: it is due to the
increase of the geocoronal \lya\ line as the spacecraft moves toward the day
side of its orbit. This effect is most severe in the last two orbits of visit~1
and all three orbits of visit~2, where the background level during the last
exposures reaches the level of \lya\ brightness in the photometric aperture.
The peak of the \lya\ emission line and is estimated to $\sim
12$~photoelectrons (e$^-$) from the median spectrum of
Fig.~\ref{fig:spectrum_2d}.

Note that the absence of a dispersing slit makes the geocoronal \lya\ emission
contributing to the flux measured on every pixels of the detector. Therefore,
the airglow contamination can be corrected for by performing the background
subtraction. These exposures where the signal is overwhelmed by the geocoronal
emission (see Fig.~\ref{fig:background}) are discarded from the photometric
time series.

The two time series obtained for visits~1 and~2 are phase-folded, using the
ephemeris derived by Wittenmyer et al.\ (2005), to yield two transit light
curves containing 26 and 15 points, respectively. The curves are plotted in
upper and middle panels of Fig.~\ref{fig:transit_light_curves}, where each
point represent the weighted mean of all individual exposures within a given
\emph{HST} orbit. We fit to all sub-exposures from each time series a simple
trapezoidal transit curve, where the geometry, i.e., the impact parameter and
the transit, ingress, and outgress durations, are obtained from accurate
photometry of the transit light curve in the optical (Knutson et al.\ 2007).
The transit depth and the out-of-transit baseline level are free to vary to fit
the data. Values for these free parameters are then obtained by minimizing a
$\chi^2$.

\section{Results}

\subsection{The size of the extended exosphere: comparison with previous measurements}

In the first transit (visit~1), we obtain a $\chi^2$ of 22.5 for $n = 24$
degrees of freedom, or a $\chi^2/n \approx 0.94$ and measure a dip of
$(9.6\pm7.0)\%$ for an out-of-transit baseline level of $301 \pm 13$~e$^-$. The
fit to the second transit (visit~2) gives a $\chi^2/n = 16.5 / 13 \approx 1.3$,
an absorption depth of $(5.3\pm10.0)\%$, consistant with visit~1, and a
baseline level of $281\pm19$~e$^-$. Note that the measured errors in the
determination of the out-of-transit baseline levels are very close to the
theoretical errors expected for a photon-noise-limited signal.

The out-of-transit baseline levels are then used to normalize the respective
time series (lower panel of Fig.~\ref{fig:transit_light_curves}). We finally
repeat the above described fitting procedure on normalized time series to
obtain a transit depth while considering data points from both visits together.
We obtain a depth of $(8.0\pm5.7)\%$, in good agreement with individual time
series. This last global fit initially yielded a $\chi^2/n = 67.6 / 40 \approx
1.7$, meaning that the propagated errors underestimate the actual dispersion of
the photometry, by a factor of $\sqrt{1.7} \approx 1.3$. The error bars of
individual exposures were thus scaled larger by a factor of $1.3$ to reflect
their true dispersion, and so that the value of $\chi^2/n$ for the global fit
is close to $1$. This is a conservative approach for the final quoted error
bars.

The transit depth in the visible is $\sim 1.6\%$ (Brown et al.\ 2001). The
observed difference between visible and \lya\ measurements is indicative of an
additional absorption centered around the \lya\ emission line. The measured
absorption is compatible with the $(5\pm2)\%$ measured by Vidal-Madjar et al.\
(2004) at an equivalent resolution with \emph{HST}/STIS. As discussed in
Vidal-Madjar et al.\ (2008), such a result obtained by measuring the absorption
on the whole unresolved \lya\ line, is in agreement with other measurements
made at a higher resolution. In fact, Vidal-Madjar et al.\ (2003) measured a
transit depth of $(15\pm4)\%$ on $\sim 1/3$ of the line, while Ben-Jaffel
(2007) reported an absorption of $(9\pm2)\%$ over $\sim 1/2$ of the line.

The absorption of the stellar flux during the transit is approximately equal to
the ratio of the planetary to stellar surfaces, $(R_p / R_\star)^2$. An
absorption of $(8.0\pm5.7)\%$ thus corresponds to the \emph{passage} of a
spherical hydrogen cloud of radius $R_{\rm H} \approx (0.28\pm0.1)$~$R_\star$,
where $R_\star = 1.12$~R$_\odot$ is the stellar radius (Knutson et al.\ 2007).
It gives $R_{\rm H} = (3.1\pm1.1)$~Jovian radii ($\mathrm{R_J}$), i.e., much
larger than the planetary radius of $R_p = (1.32\pm0.02)~\mathrm{R_J}$ measured
in the optical by Knutson et al.\ (2007).

The size of the inflated hydrogen envelope can also be compared to the size of
the planet Roche lobe to determine whether the atmosphere is evaporating. The
Roche lobe is not spherical but rather elongated in the star-planet direction
(Lecavelier des Etangs 2004). In the observational configuration of a transit,
the Roche limit of relevance should be taken perpendicular to the star-planet
axis; in this case $R_\mathrm{Roche} \approx 0.33~R_\star$ or
$3.7~\mathrm{R_J}$ (see Vidal-Madjar et al.\ 2003, 2008). Considering the size
$R_{\rm H}$ of the hydrogen exosphere and its $1$-$\sigma$ uncertainty, it is
possible that the exosphere reach and exceed the Roche limit.

In fact, we recall that the absorption is measured over the whole \lya\ line.
Measuring it over the line core, which would be permitted with a higher
spectral resolution, would give a larger absorption (Vidal-Madjar et al.\
2008), corresponding to a hydrogen envelope extended well beyond the Roche
limit. In other words, the velocity of hydrogen atoms responsible for the
absorption must largely exceed the planet escape velocity of $\sim
54$~km\,s$^{-1}$ (at the level of the dense atmosphere, far below the
exosphere). Vidal-Madjar et al.\ (2003) reported an absorption in the resolved
\lya\ line ranging from $-130$ to $+100$~km\,s$^{-1}$. Their observed
$(15\pm4)\%$ absorption over this velocity range corresponds to about
$(5.7\pm1.5)\%$ absorption of the total \lya\ line intensity (Vidal-Madjar et
al.\ 2004). The $(8.0\pm5.7)\%$ absorption of the total \lya\ line intensity
from the present ACS data set could thus even be compatible with hydrogen atoms
velocities larger than $\sim 200$~km\,s$^{-1}$.

Because of the observed absorption over the unresolved \lya\ line, and from
these independent observations only, the hydrogen upper atmosphere must either
extend beyond the Roche lobe (if the absorption occurs within the narrow core
of the \lya\ line), or the atoms velocities must exceed the escape velocity (if
the absorption occurs over the whole line or over a broad velocity range of
about $\pm 200$~km\,s$^{-1}$). Since the atmospheric escape takes place in both
cases, the present result is a new independent confirmation of the presence of
an extended hydrogen exosphere around HD\,209458b, first observed and confirmed
by Vidal-Madjar et al.\ (2003, 2004), which strengthens the atmospheric escape
scenario.

\subsection{The hydrogen tail: comparison with models}

\begin{figure}
\resizebox{\hsize}{!}{\includegraphics{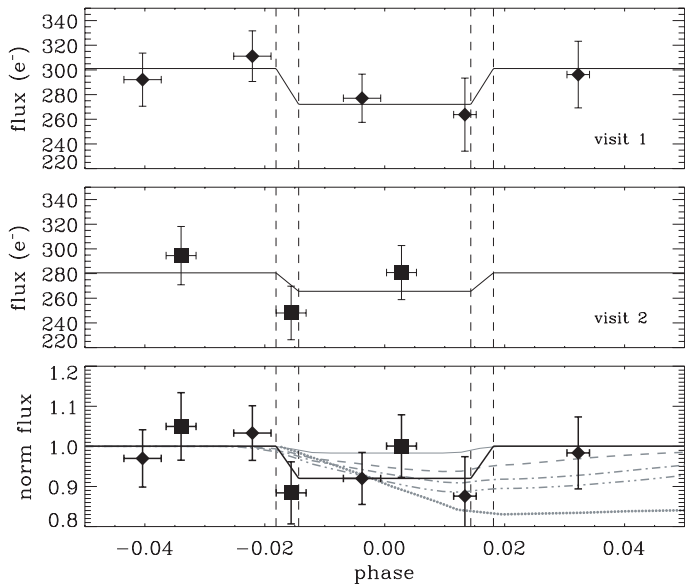}} \caption{Light curves
obtained during visits~1 (filled diamonds, upper panel) and~2 (filled squares,
middle panel). Each data point is the weighted mean of values gathered during
one \emph{HST} orbit. The phase coverage of each \emph{HST} orbit is
represented by an horizontal error bar. Trapezoidal transit curves are fitted
to each time series (plain lines). Contacts are spotted by vertical dashed
lines. Each time series is normalized by the out-of-transit flux of its best
fit. Resulting values are plotted in lower panel and fitted together with a
trapezoidal transit curve (thick black line). The transit depth seen in the
visible is indicated for comparison (thin grey line). Predicted light curves of
the transit of a hydrogen tail at low spectral resolution produced by two
different modeling approaches are also represented in lower panel: (i) assuming
the evaporation tail is ionized by the stellar EUV flux (Lecavelier des Etangs,
in preparation), for $\dot{M} = 1 \times 10^9$~g\,s$^{-1}$ (dashed line),
$\dot{M} = 1 \times 10^{10}$~g\,s$^{-1}$ (dash-dotted line), and $\dot{M} = 3
\times 10^{10}$~g\,s$^{-1}$ (dash-dot-dotted line); (ii) assuming the
evaporation tail interacts with the ionized stellar wind (Schneiter et al.\
2007), for $\dot{M} = 1 \times 10^{10}$~g\,s$^{-1}$ (dotted line).}
\label{fig:transit_light_curves}
\end{figure}

The spectral resolution of data analyzed in the present study does not allow us
to constrain the velocity of the observed hydrogen and compare it to the escape
velocity, as done by Vidal-Madjar et al.\ (2003).  A clear signature of
hydrogen escaping from the planet gravity would be the observation of an
escaping hydrogen `cometary-like' tail. Materials expelled from the planet in
the observer's direction by effects such as the stellar radiation pressure are
seen projected on the plan perpendicular to the line of sight. Because of the
almost circular revolution of the planet, the gas tail would hence appear
trailing the planet orbit (Rauer et al.\ 2000; Moutou et al.\ 2001). Recent
predictions of the observational signature of the evaporation tail were
provided by Schneiter et al.\ (2007). They assumed that the planet blows an
isotropic neutral hydrogen wind at the escape velocity, with a mass loss rate
set as a free parameter. Escaping atomic H atoms are submitted to gravitational
forces from the star and the planet, and interact with the impinging ionized
stellar wind. The interaction with the stellar wind, emitted from the star at a
rate of $5.7 \times 10^{12}$~g\,s$^{-1}$, carves the escaping atmosphere into a
comet-like tail (see their Fig.~1). When transiting the star following the
planet itself, this tail produces a transit light curve which shape is changing
depending on the assumed mass loss rate $\dot{M}$ of the planet. A simulated
light curve with $\dot{M} = 1.6 \times 10^{-16}$~M$_\odot$\,yr$^{-1} = 1 \times
10^{10}$~g\,s$^{-1}$ (model M2 in Schneiter et al.\ 2007), obtained by
integrating the flux between -320 and 200~km\,s$^{-1}$ around the center of the
\lya\ line, i.e., on the whole line, is plotted over our data points in
Fig.~\ref{fig:transit_light_curves}.

A similar evaporation tail is obtained in simulations performed within our
team, and which were used in previous studies (Vidal-Madjar et al.\ 2003).
These simulations include the effect of gravity, stellar radiation pressure --
which shapes the evaporation cloud into a comet-like tail --, and stellar EUV
ionizing flux -- which limits the lifetime of H atoms escaping from the planet
(Lecavelier des Etangs, in preparation). The simulated absorption in the light
curve was computed for $\dot{M} = 1 \times 10^9$, $1 \times 10^{10}$, and $3
\times 10^{10}$~g\,s$^{-1}$ assuming a low spectral resolution, on the whole
unresolved line, as this is the case for data treated by Vidal-Madjar et al.\
(2004) and in the present work. The value of the ionizing stellar EUV flux in
these simulations was set to the solar value.

Resulting light curves are plotted in Fig.~\ref{fig:transit_light_curves}
(lower panel). The simulated absorptions over the unresolved \lya\ line fit our
measurements correctly; however, the precision achieved in the data set does
not allow us to put strong constraints on $\dot{M}$. Indeed, the presence of an
evaporation tail is mainly constrained by data from the last \emph{HST} orbit
of visit~1. While models of an evaporation tail ionized by stellar EUV
radiation are all compatible with our measurement within $1\sigma$, the model
of Schneiter et al.\ (2007) seems to overestimate the absorption of the
hydrogen tail by $\ga 8\%$ after 4th contact. We suggest that this might be
caused either by the width of the spectral interval chosen to integrate the
\lya\ absorption, which is barely covering the whole line, or by the value of
the stellar wind mass loss rate assumed by these authors.

In addition, Schneiter et al.'s model do not include photoionization, that
would decrease the length of the neutral hydrogen tail. Furthermore, none of
the models presented here are considering charge exchanges with stellar wind
protons which might occur in the exosphere of the planet and which effect would
be to lower the escape rate needed to account for the absorption observed
(Holmstr\"om et al.\ 2008). In fact, Holmstr\"om et al.\ (2008) show that this
last mechanism could produce $\sim 10\%$ of the escape rate evaluated
considering the stellar radiation pressure and gravity (Lecavelier des Etangs
et al.\ 2004). However, because both the stellar radiation pressure and gravity
field cannot be neglected in the present situation, the charge exchange process
proposed by Holmstr\"om et al.\ is an \emph{additional effect} rather than an
alternative scenario. Hence, it could only enhance the escape rate above the
original value evaluated by Vidal-Madjar et al.\ (2003).

Nonetheless, looking back to our ACS data, more exposures obtained after 4th
contact would be needed in order to test the presence of the hydrogen tail,
better constrain $\dot{M}$ for a given stellar EUV flux, and precise the
physical processes contributing to the observed absorption.

\section{Conclusions}

We observed two transits of the extrasolar planet HD\,209458b in the stellar
\lya\ emission line with \emph{HST}/ACS. Absorption depths of $(9.6\pm7.0)\%$
and $(5.3\pm10.0)\%$ were measured some $15$-days apart. No significative
evidence for a variability in the size of the hydrogen exosphere can be
determined from these data.

The extended size of the exosphere is nevertheless confirmed and the absorption
obtained by fitting data from both transits, $\Delta F / F = (8.0\pm5.7)\%$, is
in agreement with previous measurements at similar spectral resolution
(Vidal-Madjar et al.\ 2004). Because of the observed absorption over the
unresolved \lya\ line, the hydrogen upper atmosphere must extend beyond the
Roche lobe, or equivalently the velocity of hydrogen atoms must largely exceed
the escape velocity (i.e., the \lya\ line is larger than $\sim
200$~km\,s$^{-1}$).

This study is limited by the low accuracy of photometry constrained by photon
noise. The rise of the sky background during each orbit of the spacecraft, due
to the increasing contribution of the geocoronal \lya\ emission, is an
intrinsic problem of slitless spectroscopy. Higher resolution and more
sensitive slit spectroscopy, associated with a better phase coverage of the
transit aftermath, are needed to detect \emph{deneb-el-Osiris}\footnote{The
tail (\emph{deneb} in Arab) of the planet \object{Osiris}, the nickname of
HD\,209458b.}, the hydrogen tail expected by evaporation models.

While studies of the evaporation in other extrasolar planet, like HD\,189733b,
is still going on with the \emph{HST}/ACS, high hopes are placed on the repair
of STIS and the installation of the Cosmic Origin Spectrograph (COS) during
\emph{HST} Servicing Mission 4.

\begin{acknowledgements}
We thank Jeremy Walsh for his help with the scheduling of the ACS observations,
Jeffrey Linsky for his promptness to review the manuscript, and Roger Ferlet
and Michel Mayor for their support. This work is based on observations made
with the Advanced Camera for Surveys on board the NASA/ESA \emph{Hubble Space
Telescope}, obtained at the Space Telescope Science Institute, which is
operated by the Association of Universities for Research in Astronomy, Inc.,
under NASA contract NAS~5-26555. These observations are associated with program
\#10145. DE acknowledges support of the ANR `Exoplanet Horizon 2009'.
\end{acknowledgements}

\end{document}